\pdfoutput=1
\documentclass[aps,prl,floatfix,superscriptaddress,twocolumn,reprint]{revtex4-1}
\usepackage{eurosym}
\usepackage{amsbsy}
\usepackage{latexsym,epsfig,graphicx}
\usepackage{dcolumn}
\usepackage{graphicx}
\usepackage{subfigure}
\usepackage{comment}
\usepackage{color}
\usepackage{bm}
\usepackage{mathrsfs}
\usepackage{amssymb}
\usepackage{amsthm}
\usepackage{amsfonts}
\usepackage{amsmath}
\usepackage{xspace}
\usepackage{epstopdf}
\usepackage{tabularx}
\usepackage{longtable}
\usepackage[colorlinks=true, letterpaper=true, pdfstartview=FitV, linkcolor=blue, citecolor=blue, urlcolor=blue]{hyperref}
\usepackage[normalem]{ulem}
\usepackage[version=4]{mhchem}

\newcommand{\sgn}{\text{sgn}}

\begin{document}
\title{Knots and Non-Hermitian Bloch Bands} 
\author{Haiping Hu}
\affiliation{Department of Physics and Astronomy, George Mason University, Fairfax, Virginia 22030, USA}
\affiliation{Department of Physics and Astronomy, University of Pittsburgh, Pittsburgh, Pennsylvania 15260, USA}
\author{Erhai Zhao}
\email{ezhao2@gmu.edu}
\affiliation{Department of Physics and Astronomy, George Mason University, Fairfax, Virginia 22030, USA}
\begin{abstract}
Knots have a twisted history in quantum physics. They were abandoned as failed models of atoms. Only much later was the connection between knot invariants and Wilson loops in topological quantum field theory discovered. Here we show that knots tied by the eigenenergy strings provide a complete topological classification of one-dimensional non-Hermitian (NH) Hamiltonians with separable bands. A $\mathbb{Z}_2$ knot invariant, the global biorthogonal Berry phase $Q$ as the sum of the Wilson loop eigenphases, is proved to be equal to the permutation parity of the NH bands. We show the transition between two phases characterized by distinct knots occur through exceptional points and come in two types. We further develop an algorithm to construct the corresponding tight-binding NH Hamiltonian for any desired knot, and propose a scheme to probe the knot structure via quantum quench. The theory and algorithm are demonstrated by model Hamiltonians that feature for example the Hopf link, the trefoil knot, the figure-8 knot and the Whitehead link. 
\end{abstract}
\maketitle

Extending topological band theory to non-Hermitian (NH) systems has significantly broadened and deepened our understanding about the topology of Bloch bands. NH Hamiltonians \cite{coll1,coll2,coll3,coll4,coll4,coll5,coll6,colladd1,colladd2,colladd3} are effective descriptions of a diverse set of many-body systems ranging from photonic systems with gain or loss \cite{op1,op2,op3,op4,op5,op6,op7,op8,op9,op10,op11,op12,op13,op14,op15,op16,op17,op18,op19,op20,op21,op22} to quasiparticles of finite lifetime \cite{finite1,finite2,finite3,finite4,finite5,finite6,finite7,finite8}. In contrast to Hermitian systems, NH Hamiltonians have complex eigenenergies. This unique property gives rise to a number of intricate phenomena without Hermitian counterparts including for example the exceptional point (EP), where eigenstates coalesce \cite{etopo1,etopo2,etopo3,pointtopo4,etopoadd1,etopoadd2,etopoadd3}, and the NH skin effect \cite{ne1,ne2,ne3,ne4,ne5,ne6,ne7,ne8,ne9,ne10,ne11,nhsee1,nhsee2,nhsee3,nhsee4,neadd1}, where an extensive number of eigenmodes are localized at the boundary. A synopsis of earlier NH band theory is the classification of topologically distinct NH Hamiltonians based on symmetry \cite{pclass1,pclass2,pclass3,pclass4,pclass5,pclass6} akin to the Hermitian ten-fold way \cite{hclass1,hclass2,hclass3,hclass4}. This classification scheme starts by distinguishing two types of band gaps, the line gap and point gap. While NH bands with line gaps can be continuously deformed to their Hermitian counterparts, the point-gap topology is intrinsically NH \cite{pointtopo1,pointtopo2,pointtopo3} and explains the NH skin effect.

Recently it was recognized that the NH band theory in Refs. \cite{pclass1,pclass2,pclass3,pclass4} based on the gap dichotomy is incomplete. A NH Hamiltonian may not possess a well-defined point or line gap. A more general theory only assumes separable bands \cite{fuliang}, i.e. the eigenenergies $E_j(\bm k)\neq E_l(\bm k)$ for all $j\neq l$ and crystal momentum $\bm k$. Moreover the ubiquitous twisting and braiding of complex eigenenergies give rise to new topological invariants. For example, in one dimension (1D), as $k$ is varied form 0 to $2\pi$, the eigenenergy trajectories $\{E_j(k)\}$ may form a ``braid" (see Fig. \ref{fig1} below). Two topologically distinct NH band structures (two braids) cannot be continuously deformed into each other while keeping the bands separable. Based on homotopy analysis, recent work established that the distinct topological sectors of 1D NH Hamiltonians with $N$ separable bands correspond to the conjugacy classes of the braid group $B_N$ \cite{class1,class2}. Unfortunately, homotopy theory alone does not offer an algorithm to compute the invariants directly from the Hamiltonian \cite{fnote1}. This raises the following open questions. (\textit{i}) Given a generic NH Hamiltonian, how to determine its topological invariant? (\textit{ii}) How to describe the phase transition between two topologically distinct phases? (\textit{iii}) How to design a NH Hamiltonian whose bands form a desired braid pattern?
\begin{figure}[t!]
\includegraphics[width=0.47\textwidth]{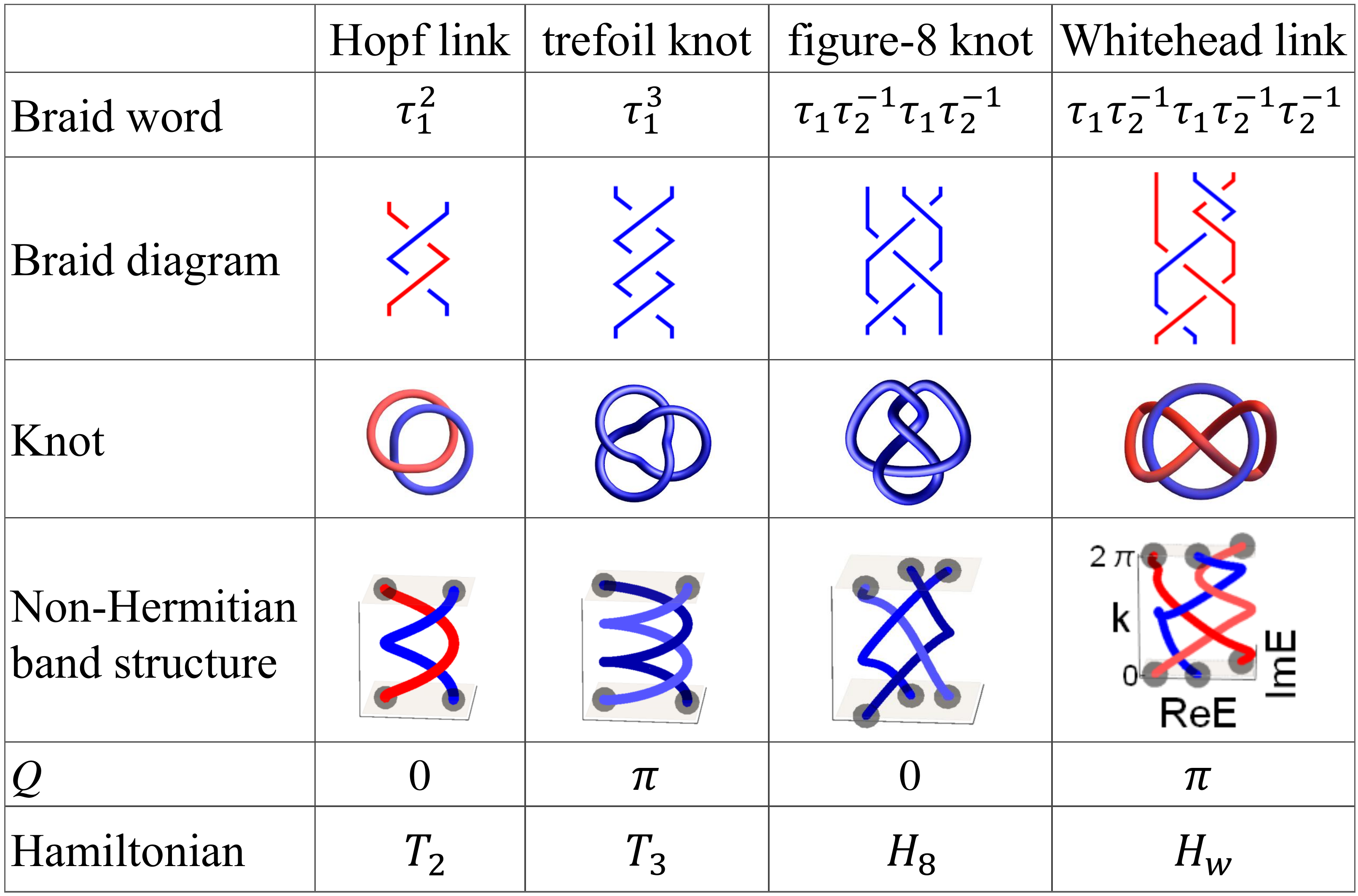}
\caption{Four examples of links/knots in 1D NH Bloch bands. Braid operator $\tau_i$ ($\tau_i^{-1}$) denotes the $i$-th string crossing over (under) the ($i$+1)-th string from left. Colors label different knot components. $Q$ is the biorthogonal Berry phase defined in Eq. (\ref{bphase}). The four knots are realized by NH Hamiltonians $T_2$, $T_3$ as defined in Eq. (\ref{model0}), $H_{8}$ and $H_{w}$ \cite{SM}, respectively. The eigenenergy strings are shown in space (Re$E$, Im$E$, $k$).}\label{fig1}
\end{figure}

In this paper, we answer these questions by developing a knot theory for NH Hamiltonians. We prove that the topology of 1D NH Hamiltonians with separable bands is fully characterized by the knots (or links) formed by the eigenenergy strings, and the topological invariants are thus knot invariants. This perspective based on knots enables us to predict two types of phase transitions accompanied by the emergence of EPs and abrupt changes in the biorthogonal Wannier centers. We also present an algorithm to design tight-binding Hamiltonians to realize arbitrary knots, and demonstrate how the knots could be revealed from quantum quench experiments and realized in electric circuits and photonic arrays. In contrast to the various knots residing in the 3D $\bm{k}$-space and formed by the zero-energy nodal lines of topological semimetals \cite{Zhong_2017,Yan_Hopf,Bi_knot,Ezawa_Hopf,Chen_Hopf,XGWan,Li_knot,jphknot,Lee_knot,knotadd1,knotadd2,knotadd3,knotadd4,knotadd5}, the knots here live in the energy-momentum space and describe the topology of the entire NH band structures.

{\color{blue}\textit{Knot classification of non-Hermitian band structures.}} Our first main result is that {\it 1D NH Hamiltonians with separable bands and no symmetry are completely classified by knots inside a solid torus}. It follows that a topological invariant of the band structure must be a {\it knot invariant}. To prove this statement, first we summarize the results of Refs. \cite{class1,class2}. A 1D NH band structure with $N$ separable bands defines a map from the Brillouin zone, a circle $S^1$, to the configuration space $X_N=(\mathrm{Conf}_N\times F_N)/S_N$. Here Conf$_N$ is the ordered $N$-tuples of complex energy eigenvalues, the quotient space $F_N=U(N)/U^N(1)$ describes the energy eigenvectors, and $S_N$ is the permutation group. Since $\pi_1(F_N)=0$, the equivalent classes of non-based map $[S^1,X_N]$ can be reduced to $[S^1,\mathrm{Conf}_N/S_N]$, and further to the conjugacy classes of the braid group $B_N=\pi_1(\mathrm{Conf}_N/S_N)$ \cite{class1,class2}. While this formal result based on homotopy theory is rigorous, the conjugacy classes of $B_N$ are hard to compute or visualize \cite{knotbook}. Here, we further relate them to knots. Notice that the braids of energy eigenvalues (constructed explicitly below) are {\it closed} due to the periodicity of the Brillouin zone, so the braid space is a solid torus. A theorem in knot theory dictates that two closed $N$-braids in $B_N$ can be smoothly deformed into each other in the solid torus  {\it iff} they are conjugate to each other \cite{knotbook}. Thus, thanks to the one-to-one correspondence between the conjugacy class of $N$-braids and knots, we reach the conclusion that {\it knots provide a natural language to classify 1D NH Bloch bands}.

It is physically intuitive to construct the knot for a given 1D NH Hamiltonian $H(k)$. The procedure is outlined as follows. The complex eigenenergies form a set $\mathscr{E}=\{E_j(k)\}$ with band index $j=1,...,N$. They are the roots of the characteristic polynomial (ChP)
\begin{eqnarray}
f(\lambda,k)=\det(\lambda-H(k))=\prod_{j=1}^N [\lambda-E_j(k)].
\end{eqnarray}
As $k$ evolves from $0$ to $2\pi$, the trajectory of $E_i(k)$ defines a {\it string} in the 3D space spanned by $(\textrm{Re}E,\textrm{Im}E,k)$. Overall $N$ such strings may tangle with each to form a braid shown in Fig. \ref{fig1}. A braid can be faithfully described by its braid diagram obtained by projecting the $N$ strings onto a chosen 2D plane parallel to the vertical $k$-axis. A braid diagram consists of a sequence of string crossings, each characterized by a {\it braid operator} $\tau_i$ in Artin's notation. For instance, when projected on plane $\textrm{Im}E=+\infty$, $\tau_i$ $(\tau_i^{-1})$ is defined by $\textrm{Re}E_i=\textrm{Re}E_{i+1}$ and $\textrm{Im}E_i<\textrm{Im}E_{i+1}~(\textrm{Im}E_i>\textrm{Im}E_{i+1})$. In other words, $\tau_i$ $(\tau_i^{-1})$ indicates that the $i$-th string crosses over (under) the $(i+1)$-th string from left. Note that two non-adjacent braid operators commute: $\tau_i\tau_j=\tau_j\tau_i$ for $|j-i|\geq2$, and $\tau_i\tau_{i+1}\tau_i=\tau_{i+1}\tau_i\tau_{i+1}$. The entire braid is then specified by its {\it braid word}, a product of braid operators, see Fig. \ref{fig1}. The set $\mathscr{E}$ is identical for $k=0$ and $k=2\pi$, so the braid is closed and becomes a knot (oriented with increasing $k$) in the $(\textrm{Re}E, \textrm{Im}E, k)$ space, which is topologically a solid torus. The end result of $k$ evolution over one period $2\pi$ is the permutation
\begin{eqnarray}\label{perm}
\sigma=\left(\begin{array}{cccc}
E_1(0) & E_2(0) & ... & E_N(0)\\
E_1(2\pi) & E_2(2\pi) & ... & E_N(2\pi)
\end{array}\right).
\end{eqnarray}
As usual, we define its parity $P(\sigma)=\pm 1$ if $\sigma$ can be expressed as even/odd number of transpositions.

The braid diagram may not be unique for a given band structure. Different choices of the projection plane yield isotopic braids related to each other by Reidemeister moves. Moreover, choosing different starting points $k_0$ for the $k$ interval $[k_0,k_0+2\pi]$ corresponds to braids within the same conjugacy class. This provides a clear understanding of why the conjugacy classes, not the elements, of $B_N$ are used for classification. These different choices however always yield the same unique knot, which is invariant under Reidemeister moves or translations along the $k$ axis. Thus using knots to describe the NH band structure is not only natural but also economical, free from the arbitrariness in braid representations. Topologically distinct NH band structures correspond to distinct knots. Fig. \ref{fig1} lists four knots, known as the Hopf link, trefoil knot, figure-8 knot, and Whitehead link. The associated braids are also shown. To avoid clutter, hereafter we will also refer to links loosely as knots.  

{\color{blue}\textit{Knot invariants}}. It follows immediately that 1D NH bands are characterized by knot invariants \cite{knotbook2,jones}. In addition to the well-known polynomial invariants \cite{SM}, here we introduce a $\mathbb{Z}_2$ topological invariant $Q$ and relate it to the parity of band permutations defined earlier. For NH Hamiltonians, the right and left eigenvectors are defined as $H(k)|\psi_{n}\rangle=E_n(k)|\psi_{n}\rangle$ and $H^{\dag}(k)|\chi_{n}\rangle=E^*_n(k)|\chi_{n}\rangle$, which satisfy the biorthogonal normalization $\langle\chi_{m}|\psi_{n}\rangle=\delta_{mn}$ \cite{biqm}. Define the non-Abelian Berry connection $A_B^{mn}=i\langle\chi_{m}|\partial_k|\psi_{n}\rangle$ and the global biorthogonal Berry phase \cite{gberry1}
\begin{eqnarray}\label{bphase}
Q=\oint_0^{2\pi}dk~\textrm{Tr}[A_B].
\end{eqnarray}
One can prove \cite{SM} that {\it $Q$ is quantized to $0$ ($\pi$) when the band permutation $\sigma$ is even (odd)},
\begin{eqnarray}
e^{iQ}=(-1)^{P(\sigma)}.
\end{eqnarray}
While $Q$ is indeed a knot invariant, due to its $\mathbb{Z}_2$ nature it only coarsely classifies knots into two groups. For example, the Hopf and figure-8 knot have the same $Q=0$, and similarly trefoil and Whitehead knot have $Q=\pi$. In Hermitian systems, Wilson loop provides a powerful characterization of band topology \cite{wannier,hoti01,fhoti}. For NH systems, we define the biorthogonal Wilson loop from the Berry connection
\begin{eqnarray}
W_B=\mathcal{P}~e^{i\oint_0^{2\pi} d k~A_B},\label{wl}
\end{eqnarray}
where $P$ denotes path ordering. Its eigenphases $\nu_n$, defined by $W_B|\mu_{n}\rangle=e^{i\nu_n}|\mu_n\rangle$, are the Wannier centers \cite{op22,wanniercenter,bwl}. It can be shown \cite{SM} that $Q=\sum_n\nu_n$.

\begin{figure}[t]
\includegraphics[width=0.47\textwidth]{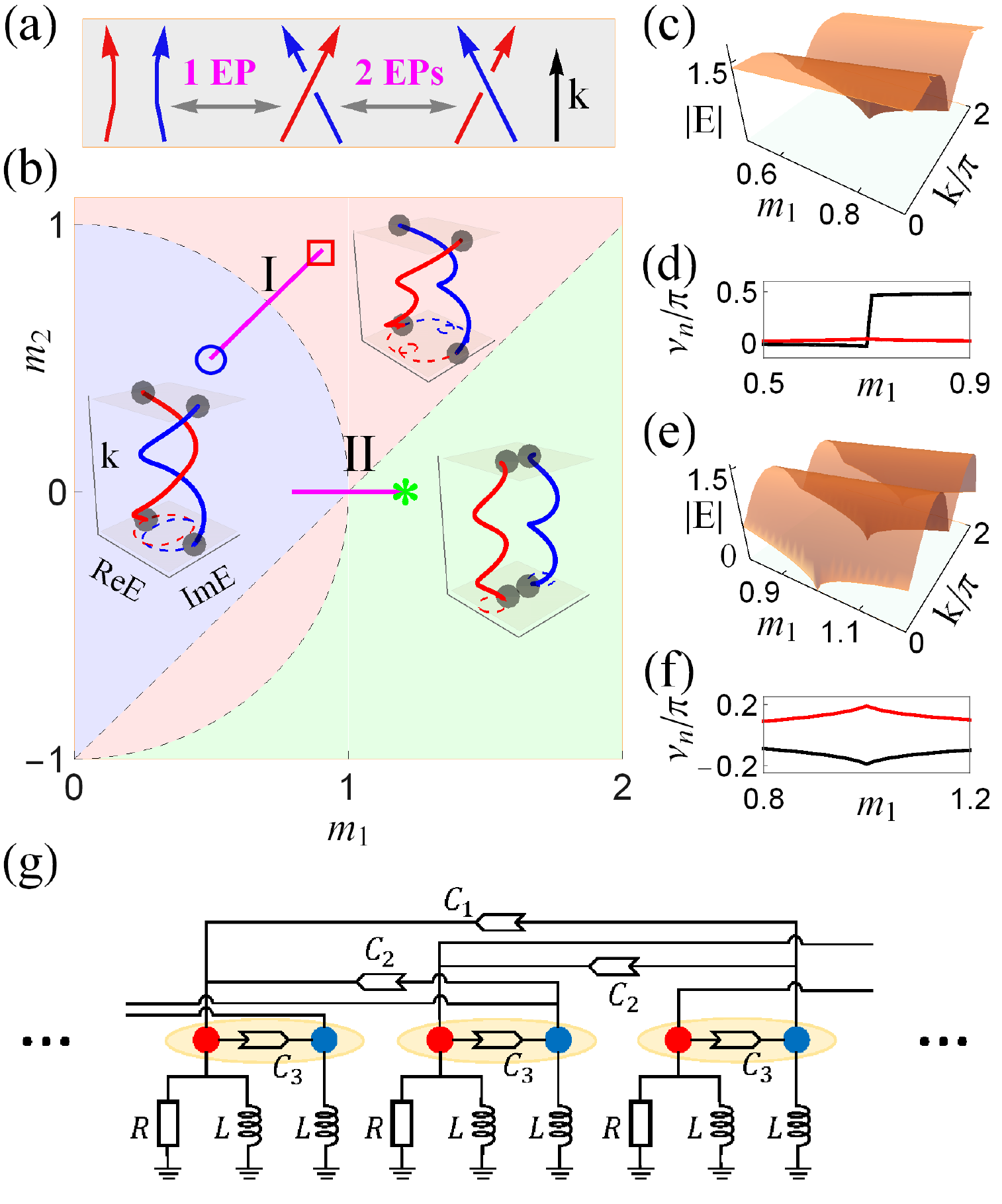}
\caption{Phase diagram and phase transitions of $H_{12}(k)$ defined in Eq. \eqref{model}. (a) Schematic of knot transitions. Type-I (type-II) transition occurs by going through one (two) EP. (b) The phase diagram of $H_{12}$ with parameters $m_1$ and $m_2$. The blue, pink, and green regions label the Hopf link ($\tau_1^2$), unknot ($\tau_1$), and unlink phase ($\tau_1^0$), respectively. In each region, a representative band structure is plotted. (c) and (e) show eigenenergy $|E(m_1,k)|$ along the cut labelled by I and II respectively in (b): an EP is visible at  $({1}/{\sqrt{2}},\pi)$ in (c), while there are two EPs at $(1,0)$ and $(1,\pi)$ in (e). (d) and (f) show the Wannier centers $\nu_n$ along the cut I and II. (g) Schematic of a periodic electric circuit that realizes $H_{12}$. The unit cell (oval) contains two ``sites", the red and blue nodes, connected by resistors $R$, inductors $L$ and negative impedance converters $C_{1,2,3}$, see \cite{SM} for details.}
\label{pd}
\end{figure}

{\color{blue}\textit{A toy model: the twister Hamiltonian.}} To illustrate different knots and their phase transitions, we introduce a simple two-band NH Hamiltonian
\begin{eqnarray}\label{model0}
T_n=\left(\begin{array}{cc}
0 & e^{i nk}\\
1 & 0
\end{array}\right),
\end{eqnarray}
where $n$ counts the number of twists of the two band strings, $E_{\pm}=\pm e^{i\frac{nk}{2}}$, as $k$ evolves from $0$ to $2\pi$. The braid word of $T_n$ is simply $\tau_1^n$. The twister \cite{fnote2} Hamiltonian $T_n$ for $n=0,1,2$ gives rise to the unlink, unknot, and Hopf link, respectively. We will use $T_n$ as building block to construct a model with two tunable parameters ($m_1$, $m_2$),
\begin{eqnarray}\label{model}
H_{12}(k)=i m_1\sigma_z+m_2T_1+T_2.
\end{eqnarray}
It has three topologically distinct phases, the Hopf link (blue region), the unlink (green), and the unknot (pink) phase, see the phase diagram in Fig. \ref{pd}(b). The phase boundaries are given by $m_1^2+m_2^2=1$ and $m_2=\pm m_1-1$. The knot topology is apparent from the two eigenenergy strings (blue and red solid lines in insets). For the unlink, the two strings do not braid, each forming a loop; for the Hopf link, the two strings braid twice, and the two loops are linked; for the unknot, the two strings braid once to form one single loop. We emphasize that all three phases here exhibit NH skin effect \cite{ne1,ne2,ne3,ne4,ne5,ne6,ne7,ne8,ne9,ne10,ne11,nhsee1,nhsee2,nhsee3,nhsee4} because projecting the knot onto the complex $E$ plane yields a band structure (dash lines) with a point gap \cite{pointtopo1,pointtopo2,pointtopo3}. Previous classification framework \cite{pclass1,pclass2,pclass3,pclass4,pclass5,pclass6} based on line/point gaps however cannot distinguish these phases or describe their phase transitions. The classification presented here based on knots is finer and complete.

{\color{blue}\textit{Phase transition through exceptional points.}} A transition between two phases characterized by different knots must occur through the crossing of the strings, i.e, through band degeneracy points. There are two kinds of band degeneracies in NH systems, the exceptional point (EP) or non-defective degeneracy point (NDP). The key difference is that EPs are defective, where the eigenvectors coalesce, leaving the Hamiltonian non-diagonalizable, while at an NDP, the eigenstates remain distinct. For a general 1D NH band with no symmetry, NDPs are unstable and will split into several EPs by small perturbations \cite{fdt}. The proof of this statement and an example can be found in \cite{SM}. Thus we are led to the conclusion that {\it a transition between phases of distinct knots is accompanied by exceptional points.}

There are two scenarios for two strings to undergo a ``knot transition" and they are sketched in Fig. \ref{pd}(a). In a type-I transition, two strings change from cross to no-cross (or vice versa) by going through an EP;  the braid word $\tau_i^{\pm 1}\rightarrow \tau_i^0$ and $Q$ also changes. One example is trefoil knot transforming to Hopf link via $\tau_1\rightarrow\tau_1^0$. A type-II transition occurs when an over-cross becomes an under-cross or vice versa,
so the braid word $\tau_i\rightarrow\tau_i^{-1}$. It is usually accompanied by two EPs, while $Q$ remains the same. 
For $H_{12}(k)$, the transition from the Hopf link to the unknot along the line $m_1=m_2$ belongs to type I and the EP is located at $(m_1,k)=(1/\sqrt{2},\pi)$, as shown in Fig. \ref{pd}(c). The transition from the Hopf link to the unlink along the $m_2=0$ line is of type II, with two EPs located at $(m_1,k)=(1,0)$ and $(1,\pi)$ as shown in Fig. \ref{pd}(d). Note that the Wannier centers undergo abrupt changes at these transitions, see Fig. \ref{pd} (d) and (f). 

{\color{blue}\textit{How to design knotty Hamiltonians.}} Beyond these simple knots, it becomes challenging to construct the tight-binding Hamiltonian $H_K(k)$ whose bands tie into certain given knot $K$. Here we outline a solution to this problem, which aids the experimental realization and probe of NH knots. The key is to find a ChP $f(\lambda,k)$ with $\lambda\in\mathbb{C}$ and $k\in[0,2\pi]$ whose roots produce the desired eigenenergy strings. Our algorithm consists of two steps \cite{SM}. In the first step, $f(\lambda,k)$ is constructed from the data of knot $K$. From the braid diagram of $K$, decompose the permutation $\sigma$ into a series of cycles $\sigma=s_1s_2...$ with $l_n$ the length of cycle $s_n$. For each cycle, standard trigonometrical parametrization \cite{knotpaper,SM} generates two real functions $F_{n}(k)$, $G_{n}(k)$. The strings in cycle $s_n$ are given by coordinates $(F_n(k^j_{n}),G_n(k^j_{n}),k)$ with $k^j_{n}=(k+2\pi j_n)/{l_n}$ and $j_n=0,...,l_n-1$. Thus the roots of the following ChP
\begin{equation}
f(\lambda,k)=\prod_{s_n}\prod_{j_n}[\lambda-F_n(k^j_{n})-iG_n(k^j_{n})] \label{braidpara}
\end{equation}
yield the desired knot $K$. The ChP obtained is a power series of $\lambda$, $f(\lambda,k)=\lambda^N+\sum_{j=0}^{N-1}\zeta_j(k) \lambda^j$, where $\zeta_j(k)$ is a Laurent series of $e^{\pm ik}$. In the second step, Hamiltonian $H_K$ is constructed from $f(\lambda,k)$ above: it is a sparse matrix \cite{SM} with the only non-zero elements being
\begin{eqnarray}
&&H_K^{i+1,i}=1,~~~~~i=1,2,...,N-1;\notag\\
&&H_K^{i,1}=-\zeta_{N-i}(k),~~i=1,2,...,N.
\end{eqnarray}
For example, applying this algorithm to braid word $\tau_1^n$ yields the twister Hamiltonian $T_n$. The NH Hamiltonians for the figure-8 knot and Whitehead link, $H_8$ and $H_w$ shown in Fig. 1, are similarly obtained. Their explicit expressions are lengthy and can be found in \cite{SM}. 
In general, more complicated knots require longer-range couplings in the tight-binding Hamiltonian. 

\begin{figure}[t]
\includegraphics[width=0.46\textwidth]{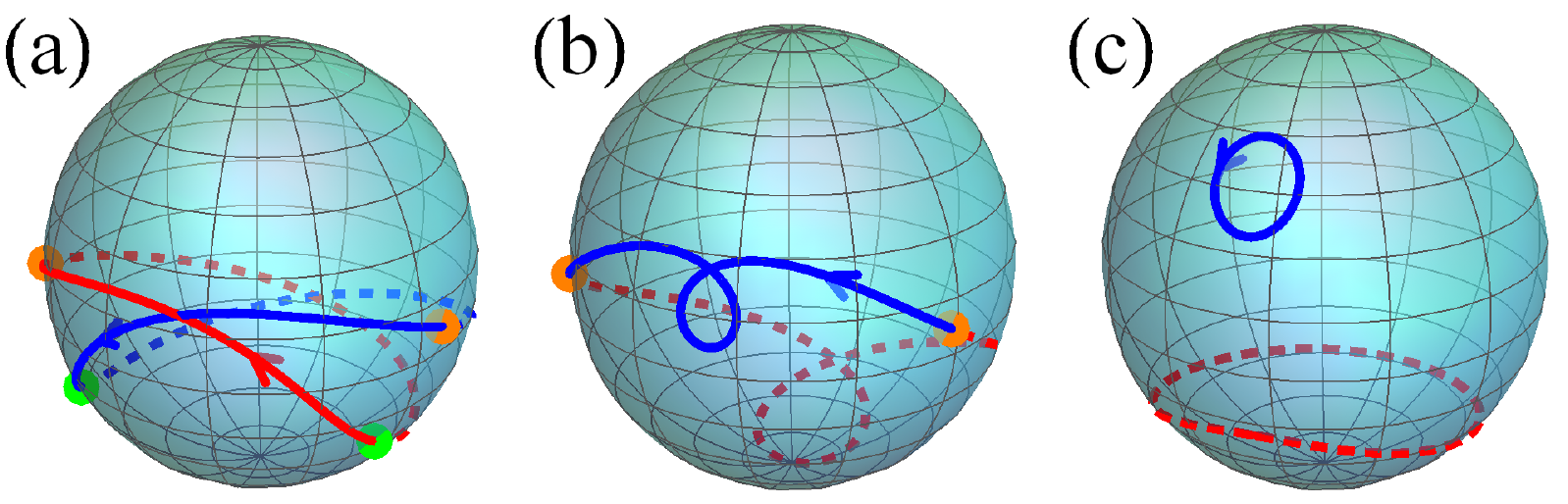}
\caption{Signatures of knots after quantum quench. The red/blue curves are the eigenvectors $|\psi_{1,2}(k)\rangle$ of $H_{12}(k)$ on the Bloch sphere. From an initial state $|\xi_0\rangle=(1,0)^T$ (north pole), the state evolves with $H_{12}(k)$ and after a long time falls into the solid line part of the eigenstates. The arrow denotes increasing $k$ from $0$ to $2\pi$, and the orange (green) dots represent the $k=0$ ($k=\pi$) mode. The parameters are (a) $m_1=m_2=0.5$, the Hopf-link phase;  (b) $m_1=m_2=0.9$, the unknot phase; and (c) $m_1=1.2, m_2=0$, the unlink phase. 
}\label{qd}
\end{figure}

{\color{blue}\textit{Experimental realization and probe of knots.}} The various proposed knots and their associated NH Hamiltonians can be realized in platforms such as photonic lattices or electric circuits \cite{zhao}. For the former, the asymmetric coupling between the sites (ring resonators) can be implemented via auxiliary microring cavities, see \cite{SM} for details. For the latter, the NH Hamiltonians can be simulated by the admittance matrix. For example, the twister Hamiltonian $H_{12}$ is simulated by the periodic circuit shown in Fig. \ref{pd}(g). It consists of resistors $R$, inductors $L$, and negative impedance converters $C_i$ that provide intra- and inter-unit cell couplings, for details see \cite{SM}.  Measurement of the admittance spectrum \cite{nhsee2,nhsee4,RLC3} yields $\{E_j(k)\}$, which provides a direct probe of the knotted band structures and the EPs.

An alternative probe of knots is through the eigenstates. As an example, consider the two-band system $H_{12}(k)$ where the eigenstates can be accessed via Bloch state tomography \cite{hhpquench,azi1,azi2,tomograph1,tomograph2}. Each of the two right eigenstates $|\psi_{1,2}(k)\rangle$ corresponds to a point on the Bloch sphere. As $k$ is varied, their trajectories trace out two curves (in red and blue) on the Bloch sphere as illustrated in Fig. \ref{qd}. For the Hopf-link phase (a), each curve is a closed loop, and they intersect twice. In the unlink phase (c), we have two closed loops that remain separated. Both phases have even permutation parity, $Q=0$. In contrast, in the unknot phase (b), the red curve joins the blue curve to form a single loop, and $Q=\pi$. It is clear from this example that the knot topology of two eigenenergy strings translates to characteristic crossing patterns of the eigenvector loops on the Bloch sphere, which can be distinguished from Bloch state tomography. The invariant $Q$ can also be read out directly.


We propose an effective way to prepare $|\psi_{1,2}(k)\rangle$ via quantum quench. From an (arbitrary) initial state $|\xi_0\rangle$ at time $t=0$, the system evolves according to $H_{12}(k)$. Let the $j$-th eigenenergy $E_j(k)=\epsilon_j-i\gamma_j$, the state at later time $t$ is $|\xi(k,t)\rangle=\sum_{j} e^{-i\epsilon_jt}e^{-\gamma_j t}\langle\chi_{j}|\xi_0\rangle|\psi_{j}\rangle$ with $\hbar=1$. Thus, after a long time, the time-evolved state will be
purified and fall into the eigenstate with smaller $\gamma_j$. Our numerical simulation of the quench dynamics verifies that starting from $|\xi_0\rangle=(1,0)^T$ (the north pole), long-time evolution will bring the state to the solid curves in Fig. \ref{qd} (the dashed curves are reached by evolution with $-H_{12}$). While the $k$-resolved tomography measurement of the quenched state does not yield the full band structure, 
different knots can be distinguished by their signatures in the eigenvectors as shown in Fig. \ref{qd} \cite{SM}.

Going beyond conjugacy classes of braid groups, we have established a knot classification of generic 1D NH Hamiltonians with separable bands: topologically distinct NH bands are described by different knots, and their transitions are through EPs. A simple model is built from $T_n$ to showcase various knots, and an algorithm is presented to construct the corresponding tight-binding Hamiltonian for any given knot. We have demonstrated 
how these knots can be experimentally realized and probed. Other physical consequences of the knotted bands, including the relation between the NH knots and skin effect will be left for further study \cite{SM}.
 An important open problem is to extend the analysis to higher dimensions and other symmetry classes, where the interplay of band braiding, eigenstate topology, and symmetries gives rise to rich unexplored phenomena, e.g., torsion invariants \cite{class1,class2}. 

\begin{acknowledgments}
This work is supported by AFOSR Grant No. FA9550-16-1-0006 and NSF Grant No. PHY-1707484.
\end{acknowledgments}

\clearpage
\onecolumngrid
\appendix
\section{Supplementary Materials}

In this supplementary materials, we provide technical details for
\begin{itemize}
\item (I) Jones polynomial as knot invariant
\item (II) Proof of the relations between the global biorthogonal Berry phase, band permutation and Wilson loop
\item (III) Stability of band touching points in a knot transition
\item (IV) The algorithm of constructing a tight-binding Hamiltonian $H_K(k)$ associated with a given knot $K$ and explicit examples of figure-8 knot and Whitehead link
\item (V) Revealing knots from quench dynamics
\item (VI) Proposals for experimental realization of Hamiltonian $H_{12}$ in electric circuits and photonic arrays
\item (VII) Non-Hermitian knot and skin effect.
\end{itemize}

\subsection{(I) Jones polynomial as knot invariant}
In the main text, we have demonstrated that the knot theory provides an intuitive characterization of one-dimensional (1D) non-Hermitian (NH) band topology. As a direct consequence, the knot classification enables characterizations of NH bands using the well-established knot invariants \cite{knotbook2}. A widely used one is the Jones polynomial \cite{jones} $V_K(q)$. By assigning a Laurent polynomial in the variable $q^{\frac{1}{2}}$ with integer coefficients, the Jones polynomial is a powerful knot invariant for discerning inequivalent oriented knots. The Jones polynomial can be easily calculated from the skein relation \cite{knotbook,knotbook2}:
\begin{eqnarray}
q^{-1}V_{K_{+}}-qV_{K_{-}}+(q^{-1/2}-q^{1/2})V_{K_{0}}=0.
\end{eqnarray}
Here $K_{0}$, $K_{+}$, $K_{-}$ refer respectively to the three oriented knots which only differ in a small region containing a projected crossing, as shown from left to right in Fig. 2(a) of the main text. We neglect crossings joined by more than two band strings in the braid diagram as they can be avoided by properly choosing the projection plane. The Jones polynomial for $N$ totally unlinked strings are known to be
\begin{eqnarray}
V_{O}(q)=(-q^{-1/2}-q^{1/2})^{N-1}.
\end{eqnarray}
Starting from the $N$ trivial NH bands without any braiding (i.e., $N$ unlinked strings), one can get the Jones polynomials for any other separable NH bands by the skein relation, through a series of strand crossings. For example, the skein relation which relates the trefoil knot (braid word $\tau_1^3$), unknot (braid word $\tau_1$, Jones polynomial $V(q)=1$) and Hopf link ((braid word $\tau_1^2$) reads
\begin{eqnarray}\label{skein1}
q^{-1}V_{trefoil}-q+(q^{-1/2}-q^{1/2})V_{Hopf}=0;
\end{eqnarray}
The skein relation which relates the Hopf link, unlink (two trivial strings, no braiding), and unknot reads
\begin{eqnarray}\label{skein2}
q^{-1}V_{Hopf}+q(q^{-1/2}+q^{1/2})+(q^{-1/2}-q^{1/2})=0.
\end{eqnarray}
From Eq. (\ref{skein1}) and Eq. (\ref{skein2}), we can solve the Jones polynomials for the Hopf link and trefoil knot
\begin{eqnarray}
V_{Hopf}(q)=-q^{5/2}-q^{1/2};~~~V_{trefoil}(q)=q+q^3-q^4.
\end{eqnarray}
By iteratively using the skein relation, one can get the Jones polynomials for the other knots in Table I in the main text, which are summarized below:
\begin{eqnarray}
\textrm{figure-8 knot:} ~~~V(q)&=&q^{-2}-q^{-1}+1-q+q^2;\notag\\
\textrm{Whitehead link:}~~~V(q)&=&-q^{-3/2}+q^{-1/2}-2q^{1/2}+q^{3/2}-2q^{5/2}+q^{7/2}.
\end{eqnarray}

\subsection{(II) Relation between biorthogonal Berry phase, band permutation and Wilson loop}
For a NH Hamiltonian $H(k)$, its right and left eigenvectors are defined as
\begin{eqnarray}
H(k)|\psi_{n}\rangle=E_n(k)|\psi_{n}\rangle,~H^{\dag}(k)|\chi_{n}\rangle=E^*_n(k)|\chi_{n}\rangle.
\end{eqnarray}
The two types of eigenvectors satisfy the biorthogonal normalization \cite{biqm} $\langle\chi_{m}|\psi_{n}\rangle=\delta_{mn}$. The global biorthogonal Berry phase is defined as $Q=\oint_0^{2\pi}dk~\textrm{Tr}[A_B]$. Here $A_B$ is the non-Abelian Berry connection matrix, with its $(m,n)$-element $A_B^{mn}=i\langle\chi_{m}|\partial_k|\psi_{n}\rangle$. $Q$ is defined modulo $2\pi$. In fact, a gauge transformation (note the biorthogonal normalization should be imposed)
\begin{eqnarray}
|\psi_n\rangle\rightarrow e^{-i\phi(k)}|\psi_n\rangle,~~\langle\chi_n|\rightarrow e^{i\phi(k)}\langle\chi_n|
\end{eqnarray}
brings $A_B^{mn}$ to $\tilde{A}_{B}^{mn}=A_B^{mn}+\partial_k\phi(k)$. $\phi(k)$ is a continuous single-valued function on $k\in[0,2\pi]$ satisfying $\phi(k=0)=\phi(k=2\pi)$. The gauge transformation takes $Q$ to $\tilde{Q}=Q+2p\pi~(p\in\mathbb{Z})$. Using all the $N$ right eigenvectors, we define an $N\times N$ matrix $\Psi=(|\psi_1\rangle,|\psi_2\rangle,...,|\psi_N\rangle)$. The global biorthogonal Berry phase is recast into
\begin{eqnarray}\label{vectorb}
Q=i\oint_0^{2\pi}dk~\textrm{Tr}[\Psi^{-1}\partial_k\Psi]=i\oint_0^{2\pi}dk~\partial_k\textrm{Tr}[\log\Psi]=i\log\frac{\det[\Psi(k=2\pi)]}{\det[\Psi(k=0)]}.
\end{eqnarray}
The periodicity of Hamiltonian $H(k)=H(k+2\pi)$ dictates that the whole eigenvector set to be identical at $k=0$ and $k=2\pi$. However due to band braiding, each eigenvector $|\psi_j\rangle$ does not necessarily return to itself by evolving from $k=0$ to $k=2\pi$. The braiding is labeled by the band permutation $\sigma$ (see Eq. (2) in the main text). It is clear from Eq. (\ref{vectorb}) that if the permutation is even, $\det\Psi(k=2\pi)=\det\Psi(k=0)$, $Q=0$; if the permutation is odd, $\det\Psi(k=2\pi)=-\det\Psi(k=0)$, $Q=\pi$. Hence $Q$ relates to the parity of band permutations through
\begin{eqnarray}\label{parity}
(-1)^{P(\sigma)}=e^{iQ}.
\end{eqnarray} 

Next we turn to the biorthogonal Wilson loop $W_B$ (see its definition in Eq. (5) in the main text). In discretized form, $W_B$ is expanded as
\begin{eqnarray}
W_B=\lim_{M\rightarrow\infty}W_b(k_{M-1})W_b(k_{M-2})...W_b(k_{1})W_b(k_0).
\end{eqnarray}
Here $k_j=\frac{2\pi}{M}j$, $\Delta k=\frac{2\pi}{M}$, and $W_b^{mn}(k_j)=\langle\chi_{m}(k_j+\Delta k)|\psi_{n}(k_j)\rangle$. By diagonalizing $W_B$, i.e., $W_B|\mu_{n}\rangle=e^{i\nu_n}|\mu_n\rangle$, we get $N$ Wannier centers $\nu_n$ $(1\leq n\leq N)$. The total Wannier center can be calculated as
\begin{eqnarray}
\lim_{M\rightarrow\infty}\sum_{n=1}^N\nu_n&=&\lim_{M\rightarrow\infty}-i\textrm{Tr}\log[W_B]\notag\\
&=&\lim_{M\rightarrow\infty}-i\log\det[W_B]\notag\\
&=&\lim_{M\rightarrow\infty}-i\sum_{j=0}^{M-1}\log\det[W_b(k_j)]\notag\\
&=&\lim_{M\rightarrow\infty}-i\sum_{j=0}^{M-1}\sum_{n=1}^N\log[\langle\chi_{n}(k_j+\Delta k)|\psi_{n}(k_j)\rangle]\notag\\
&=&\oint_0^{2\pi}dk~\textrm{Tr}[A_B]=Q.
\end{eqnarray}
We illustrate the above relations using the twister model $T_n$ (see Eq. (6) in the main text). The two eigenbands of $T_n$ and $T_n^{\dag}$ are $E_{\pm}=\pm e^{\frac{ink}{2}}$ and $E^{*}_{\pm}=\pm e^{-\frac{ink}{2}}$, with their corresponding right and left eigenvectors:
\begin{eqnarray}
|\psi_{\pm}\rangle=\frac{1}{\sqrt{2}}\left(\begin{array}{c}
e^{\frac{ink}{2}}\\
\pm1
\end{array}
\right);~~~|\chi_{\pm}\rangle=\frac{1}{\sqrt{2}}\left(\begin{array}{c}
e^{\frac{ink}{2}}\\
\pm1
\end{array}
\right).
\end{eqnarray} 
Obviously $|\psi_{\pm}\rangle$ and $|\chi_{\pm}\rangle$ satisfy the biorthogonal normalization relation. The Berry connection is
\begin{eqnarray}
i\langle\chi_{\pm}|\partial_k|\psi_{\pm}\rangle=-\frac{n}{4},
\end{eqnarray}
yielding $Q=-n\pi$. The Wannier centers are $\nu_+=\pi n$, $\nu_-=0$. For Hamiltonian $T_n$, $n$ labels the number of braidings between the two eigenbands $E_{\pm}$ by evolving $k$ from $0$ to $2\pi$. The simplest cases of $n=0,1,2,3$ correspond to unlink, unknot, Hopf link, and trefoil knot, respectively. If $n$ is even, the two bands exchange even times and the permutation $\sigma$ is even; If $n$ is odd, the two bands exchange odd times and the permutation $\sigma$ is odd. Hence Eq. (\ref{parity}) is verified.

\subsection{(III) Stability of band touching points in a knot transition}
Here we provide further evidence to support the claim that the transitions between different knotted phases are through exceprional points (EPs). While band touchings can be through either EPs or non-defective points (NDPs), only EPs are stable in 2D parameter space.

We consider a 1D NH Hamiltonian $H(m,k)$ with an external parameter $m$, $k$ is the lattice momentum. The band touchings occur within the 2D $(m,k)$ space. In the following, we focus on generic band touchings between two NH bands because band touchings shared simultaneously by three or more bands require fine tuning. The key difference between the two types of band touchings is as follows: at an EP, the two eigenvectors coalescence, while at an NDP, the two eigenvectors are linearly independent. Let us suppose the following generic form of the Hamiltonian
\begin{eqnarray}
H(m,k)=h_0(m,k)\sigma_0+h_x(m,k)\sigma_x+h_y(m,k)\sigma_y+h_z(m,k)\sigma_z
\end{eqnarray}
to describe the two-fold degeneracy. Here $\sigma_0$ is identity matrix, $\sigma_{x,y,z}$ are Pauli matrices. We will neglect the first term as it will not change the degeneracy. $h_x, h_y$ and $h_z$ are complex functions of $m$ and $k$. The spectra of $H(m,k)$ are
\begin{eqnarray}
E_{\pm}=h_0\pm\sqrt{h_x^2+h_y^2+h_z^2}.
\end{eqnarray}
At an EP or NDP, the Hamiltonian can be unitarily transformed to 
\begin{eqnarray}
\left(\begin{array}{cc}
0 & a\\
0 & 0
\end{array}\right)~~ \textrm{or} ~~\left(\begin{array}{cc}
0 & 0\\
0 & 0
\end{array}\right).
\end{eqnarray}
For the EP, $h_x^2+h_y^2+h_z^2=0$ is the sufficient condition for band degeneracy. While for the NDP, all the components of $H(m,k)$ must take $h_x=h_y=h_z=0$. To see this, we note the two eigenvectors are independent at the NDP. They form a complete basis there (denote as $V$) and  $H=V\left(\begin{array}{cc}
0 & 0\\
0 & 0
\end{array}\right)V^{\dag}=0$. It follows the existence criterion of EP: $h_x^2+h_y^2+h_z^2=0$ gives two conditions ($h_x^2+h_y^2+h_z^2$ is a complex function); while the existence of NDP requires six conditions. Hence in the 2D $(m,k)$ space, only the EP is stable; while the NDP is unstable and its existence requires fine tuning.

Here we give an example to show the instability of an NDP. We consider the following two-band model
\begin{eqnarray}
H(m,k)=\left(\begin{array}{cc}
0 & \sin k+i m\\
\sin k+i m & 0
\end{array}\right).\end{eqnarray}
There are two NDPs, located at $(m,k)=(0,0)$ and $(0,\pi)$, respectively. The two NDPs are unstable. For example, if we add an infinitesimally small perturbation $\delta \sigma_z$, the band touchings are now determined by $-\delta^2-\sin^2 k+m^2-2im\sin k=0$. Due to the perturbation, there are four band touchings at $(m,k)=(\pm\delta,0)$ and $(\pm\delta,\pi)$, coming from the splitting of the original NDP at $(0,0)$ and $(0,\pi)$, respectively.\\

\subsection{(IV) Construction of tight-binding Hamiltonian $H_K(k)$ associated with a given knot $K$}

In the main text, we have outlined the algorithm to generate a NH Hamiltonian $H_K(k)$ corresponding to an arbitrary knot $K$. The algorithm is decomposed into two steps. The first step is to find a characteristic polynomial (ChP) $f(\lambda,k)~(\lambda\in\mathbb{C},k\in[0,2\pi])$ such that its roots form the desired knot $K$. Note that $f(\lambda,k)$ is a complex-valued polynomial and contains three real variables $\mathrm{Re}\lambda$, $\mathrm{Im}\lambda$, $k$. Hence its roots can be regarded as the intersection of the two surface determined by $\mathrm{Re}f=0$ and  $\mathrm{Im}f=0$. The second step is to construct the tight-binding Hamiltonian $H_K(k)$ with $f(\lambda,k)$ as its ChP. In our algorithm, the ChP is a power series of $\lambda$ and Laurent series of $e^{\pm ik}$. Here we detail the above steps and showcase the procedures with the figure-8 knot and Whitehead link.\\

\noindent\underline{\textit{\textbf{Step-1}}} In the first step, we need to parameterize the knot $K$, which is presented by a braid diagram $B_K$ \cite{knotpaper}. Note that while $B_K$ is not unique, different choices of $B_K$ either correspond to braids related by Reidemeister moves or braids inside the same conjugacy class. We choose one specific diagram and plot it on the $xz$ plane (see Fig. 1 in the main text). The vertical $z$-axis denotes $k$ direction. For simplicity, the diagram $B_K$ is plotted in a way where the crossings are evenly distributed along the $z$-axis. Suppose there are $c[K]$ crossings in total. They are located at
\begin{eqnarray}
k_m=\frac{\pi}{c[K]}(2m-1),~~~m=1,2,...,c[K].
\end{eqnarray} 
In the two-dimensional (2D) braid-diagram presentation, each strand of $B_K$ is a piecewise linear function of $k$. $B_K$ represents for $N$ strings in 3D, with trajectories $(F_j(k),G_j(k),k),~j=1,2,...,N$. Here $F_j(k)$ and $G_j(k)$ are real functions of $k$. Our task is to obtain $F_j(k)$ and $G_j(k)$ from $B_K$. Due to string braidings, $F_j(k)$ and $G_j(k)$ are in general not $2\pi$-periodic. However $F_j(2\pi)=F_{j'}(0)$ and $G_j(2\pi)=G_{j'}(0)$ always hold for some $1\leq j'\leq N$ ($k=0$ and $k=2\pi$ are identical). This motivates us to obtain $F_j(k)$ (same for $G_j(k)$) from a parent function, where each $F_j(k)$ corresponds to a piece of the parent function. 

The $N$ strings are associated with an element $\sigma$ of the permutation group $S_N$, as defined in Eq. (2) in the main text. In group theory, $\sigma$ can be decomposed into a sequence of cycles $\sigma=s_1s_2...$. We denote $\mathscr{C}_K=\{s_1,s_2,...\}$ as the set of cycles, which gives all the link components of the closure of $B_K$ (or knot $K$). For a given cycle $s_n\in\mathscr{C}_K$, we denote $l_n$ as its length. Inside each cycle $s_n$, we rearrange its $l_n$ string indices to be from $0$ to $l_n-1$ such that the end point of $j_n$-th string at $k=2\pi$ is the starting point of the $(j_n+1)$-th string at $k=0$ for every $0\leq j_n\leq l_n-1$. Using the above notations, any string of the diagram $B_K$ is specified by a pair of indices $(s_n,j_n)$, with $s_n\in\mathscr{C}_K,~j_n=0,1,...,l_n-1$. We assign two continuous real functions $F_n(k)$ and $G_n(k)$ as parent functions, which are $2\pi$-periodic, for each link component $s_n$. The $j_n$-th string inside $s_n$ takes
\begin{eqnarray}
F_{j_n}(k)=F_n(k_n^j),~G_{j_n}(k)=G_n(k_n^j),~~~\textrm{with}~~j_n=0,1,...,l_n-1,~k\in[0,2\pi],
\end{eqnarray}
where $k_n^j=(k+2\pi j_n)/l_n$. Next we demonstrate how to obtain $F_n(k)$ and $G_n(k)$ of each cycle $s_n$ from the trigonometric interpolation of the diagram $B_K$. To get $F_n(k)$, we first neglect the crossings of $B_K$ while encode the crossing information into $G_n(k)$. For cycle $s_n$, we define a piecewise linear function $L_n(k)$ on $k\in[0,2\pi]$:
\begin{eqnarray}
L_n(k_n^j)=B_K(k)|_{s_n,j_n}, ~~~\textrm{with}~~j_n=0,1,...,l_n-1;~s_n\in\mathscr{C}_K.
\end{eqnarray}
Here $B_K(k)|_{s_n,j_n}$ denotes the $(s_n,j_n)$-th string of $B_K(k)$. The trigonometric interpolation of $F_n(k)$ is through the following $c[K]l_n$ points located at
\begin{eqnarray}
(\frac{k_m}{l_n}-\frac{\pi}{c[K]l_n},L_n(\frac{k_m}{l_n}-\frac{\pi}{c[K]l_n})),~~~m=1,2,...,c[K]l_n.
\end{eqnarray}
The interpolation data is evenly distributed along $k$ direction, hence the interpolation is the Fourier transformation:
\begin{eqnarray}
F_n(k)&=&\sum_{m=-c[K]l_n/2+1}^{c[K]l_n/2-1}a_m e^{i m k}+a_{\frac{c[K]l_n}{2}}\cos \frac{c[K]l_n}{2}k,~~~~\textrm{if}~~c[K]l_n=even,\notag\\
F_n(k)&=&\sum_{m=-c[K]l_n/2+1/2}^{c[K]l_n/2-1/2}a_m e^{i m k}, ~~~~\textrm{if}~~c[K]l_n=odd,
\end{eqnarray}
where the Fourier coefficients are
\begin{eqnarray}
a_m=\frac{1}{c[K]l_n}\sum_{n=0}^{c[K]l_n-1}L_n(\frac{k_n}{l_n}-\frac{\pi}{c[K]l_n})e^{-i(\frac{k_n}{l_n}-\frac{\pi}{c[K]l_n})m}.
\end{eqnarray}

Having obtained $F_n(k)$ for all cycles $s_n\in\mathscr{C}_K$, the next step is to determine $G_n(k)$ from $F_n(k)$ by incorporating the string crossings. Each crossing is assigned a $+$ or $-$ sign from the braid diagram $B_K$. We denote the $z$-coordinate of the crossing point as $k_p$, which are the solutions of 
\begin{eqnarray}
F_{n}(\frac{k_p+2\pi j_{n}}{l_{n}})=F_{n'}(\frac{k_p+2\pi j_{n'}}{l_{n'}}),~~~\textrm{for all}~s_n,s_{n'}\in\mathscr{C}_K;~j_{n}=0,1,...,l_{n}-1;~ j_{n'}=0,1,...,l_{n'}-1.
\end{eqnarray}
The interpolation data for $G_n(k)$ is chosen as $(\frac{k_p+2\pi j_n}{l_n},\sgn(k_p))$. Here $\sgn(k_p)=\pm 1$ if the crossing at $k_p$ is an under/over crossings in the diagram $B_K$. Compared to $F_n(k)$, usually the interpolation data of $G_n(k)$ is not evenly distributed along $k$ direction. Suppose there are $c[F_n]$ crossing points (including both crossings with itself and other component $n'\in\mathscr{C}_K$). Formally, we set the interpolation function as
\begin{eqnarray}
G_n(k)&=&\sum_{m=-c[F_n]/2+1}^{c[F_n]/2-1}b_m e^{imk}+b_{\frac{c[F_n]}{2}}\cos \frac{c[F_n]}{2}k,~~\textrm{if}~~c[F_n]=even,\notag\\
G_n(k)&=&\sum_{m=-c[F_n]/2+1/2}^{c[F_n]/2-1/2}b_m e^{imk},~~~\textrm{if}~~c[F_n]=odd,
\end{eqnarray}
The interpolation coefficient $b_m$ can be obtained by solving a matrix equation using the above interpolation data.

Through the above procedures, the strings of knot $K$ are parameterized by $(F_n(k_n^j),G_n(k_n^j),k)$, with $s_n\in\mathscr{C}_K$, $j_n=0,1,...,l_n-1$. The desired ChP with such $N$ strings as its roots are 
\begin{eqnarray}\label{che}
f(\lambda,k)=\prod_{s_n\in\mathscr{C}_K}\prod_{j_n=0}^{l_n-1}[\lambda-F_n(k_n^j)-iG_n(k_n^j)].
\end{eqnarray}\\

\noindent\underline{\textit{\textbf{Step-2}}} 
The second step is to generate an $N$ by $N$ NH Hamiltonian $H_K(k)$, with $f(\lambda,k)$ as its ChP. To this end, we expand $f(\lambda,k)$ in the powers of $\lambda$,
\begin{eqnarray}
f(\lambda,k)=\lambda^N+\sum_{j=0}^{N-1}\zeta_j(k) \lambda^j,
\end{eqnarray}
where $\zeta_j(k)~(j=0,1,...,N-1)$ is a Laurent series of $e^{\pm ik}$. There are many different choices of $H_K(k)$, corresponding to the same ChP. In the main text, we have set $H_K(k)$ as the following simple form:
\begin{eqnarray}\label{hk}
H_K(k)=\left(\begin{array}{cccccc}
-\zeta_{N-1}(k) & -\zeta_{N-2}(k) &~~~ ...~~~&~~~ ...~~~& -\zeta_{1}(k) & -\zeta_{0}(k)\\
1 & 0 & 0 & ... & ... & 0\\
0 & 1 & 0 & 0 &...& ...\\
... & 0& 1 & 0 & 0 & ...\\
... & ... & 0& 1 & 0 & 0\\
... & ...& ... & 0 & 1 & 0
\end{array}\right).
\end{eqnarray}
It is easy to check $\det(\lambda-H_K(k))=f(\lambda,k)$.\\

\begin{figure}[h!]
\centering
\includegraphics[width=5.5in]{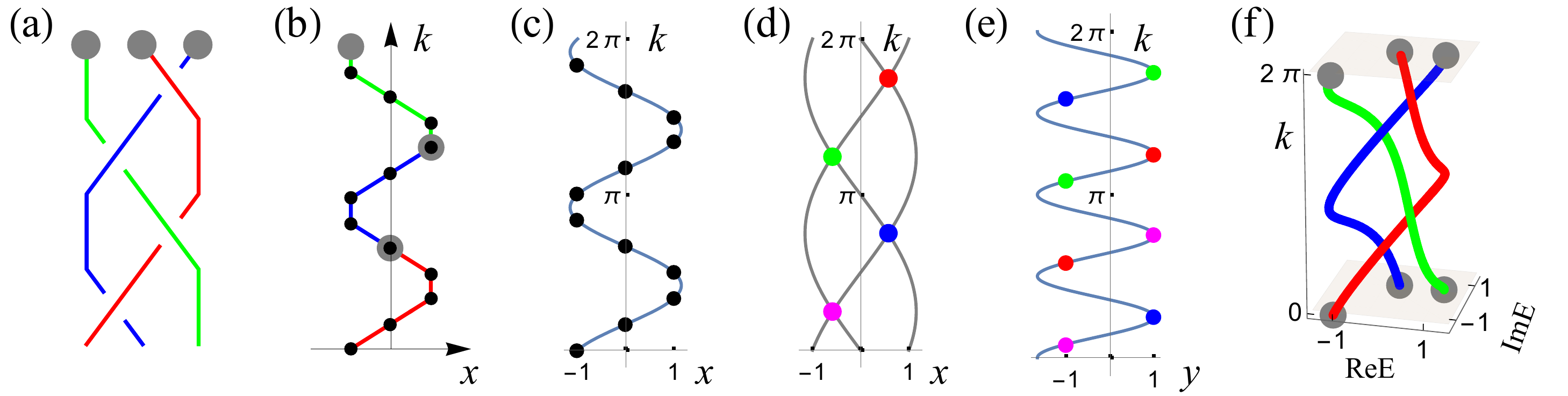}
\caption{Illustration of the algorithm of constructing tight-binding Hamiltonian $H_8(k)$ for figure-8 knot. (a) Braid diagram $B_K$ with braid word: $\tau_1\tau_2^{-1}\tau_1\tau_2^{-1}$. (b) Interpolation data (black dots) for $F_1(k)$ on the $xz$ plane. (c) Trigonometric interpolation function $F_1(k)$. (d) Crossings of the three branches $F_1(\frac{k}{3})$, $F_1(\frac{k+2\pi}{3})$, $F_1(\frac{k+4\pi}{3})$. (e) Interpolation data for $G_1(k)$ on the $yz$ plane. $\pm 1$ indicates an under/over crossing of the braid diagram. (f) Energy bands of the constructed NH Hamiltonian $H_8(k)$ defined in Eq. (\ref{hkf8}) in the 3D $(\mathrm{Re}E,\mathrm{Im}E,k)$ space, which form a figure-8 knot isotopic to the diagram $B_K$ in (a).}
\label{f8}
\end{figure}

\noindent\underline{\textit{\textbf{Example 1: figure-8 knot}}} We showcase the above procedures by explicitly working out the figure-8 knot. The braid diagram $B_K$ is depicted in Fig. \ref{f8}(a) (see also Fig. 1 in the main text), with braid word $\tau_1\tau_2^{-1}\tau_1\tau_2^{-1}$ and crossing number $c[K]=4$. By connecting the two ends at $k=0$ and $k=2\pi$, $B_K$ represents for the figure-8 knot. The string permutation of $B_K$ is $\sigma=\left(\begin{array}{cccc}
1 & 2 & 3\\
2 & 3 & 1
\end{array}\right)$. There is only one cycle $s_1=(231)$ in $\sigma$, with length $l_1=3$. 

To parameterize the braid, we first identify the data points of trigonometric interpolation for the parent functions $F_1(k)$, $G_1(k)$. Let's start with $F_1(k)$ and pick $c[K]l_1=12$ evenly distributed points along $k$ direction as shown in Fig. \ref{f8}(b). Their coordinates on the $xz$ plane are $(0,-1)$, $(\frac{\pi}{6},0)$, $(\frac{2\pi}{6},1)$, $(\frac{3\pi}{6},1)$, $(\frac{4\pi}{6},0)$, $(\frac{5\pi}{6},-1)$, $(\pi,-1)$, $(\frac{7\pi}{6},0)$, $(\frac{8\pi}{6},1)$, $(\frac{9\pi}{6},1)$, $(\frac{10\pi}{6},0)$, $(\frac{11\pi}{6},-1)$. A discrete Fourier transformation yields 
\begin{eqnarray}
F_1(k)=-\cos 2k+0.58\sin 2k,
\end{eqnarray}
as depicted in Fig. \ref{f8}(c). To find the interpolation data for $G_1(k)$, we solve all the crossings of the three strings $F_1(\frac{k}{3})$, $F_1(\frac{k+2\pi}{3})$, $F_1(\frac{k+4\pi}{3})$ inside $[0,2\pi]$. The solutions of $F_1(\frac{k}{3})=F_1(\frac{k+2\pi}{3})$ are $k_p=0.7824,5.4948$. The solution of $F_1(\frac{k}{3})=F_1(\frac{k+4\pi}{3})$ is $k_p=2.3532$. The solution of $F_1(\frac{k+2\pi}{3})=F_1(\frac{k+4\pi}{3})$ is $k_p=3.9240$. We plot all the crossing points in Fig. \ref{f8}(d). The interpolation data for $G_1(k)$ on the $yz$ plane are $(\frac{0.7824}{3},-1)$, $(\frac{2.3532}{3},1)$, $(\frac{5.4948}{3},-1)$, $(\frac{0.7824+2\pi}{3},1)$, $(\frac{3.9240+2\pi}{3},-1)$, $(\frac{5.4948+2\pi}{3},1)$, $(\frac{2.3532+4\pi}{3},-1)$, $(\frac{3.9240+4\pi}{3},1)$. Here $\pm 1$ indicates an under/over crossing of the braid diagram [see Fig. \ref{f8}(a)], respectively. The trigonometric interpolation reads
\begin{eqnarray}
G_1(k)=-0.33-1.33\cos 4k,
\end{eqnarray}
which is plotted in Fig. \ref{f8}(e). According to Eq. (\ref{che}), the ChP is
\begin{eqnarray}
f(\lambda,k)&=&\prod_{j=0}^2[\lambda-F_1(\frac{k+2\pi j}{3})-iG_1(\frac{k+2\pi j}{3})]\notag\\
&=&\lambda^3+\zeta_2(k)\lambda^2+\zeta_1(k)\lambda+\zeta_0(k).\notag\\
\zeta_2(k)&=&i;~~\zeta_1(k)=-2i\cos 2k+1.16i\sin 2k;~~\zeta_0(k)=0.73i-0.67\cos 2k-0.59i\cos 4k-1.54\sin 2k.
\end{eqnarray} 
The NH Hamiltonian $H_{8}(k)$ follows from Eq. (\ref{hk}):
\begin{eqnarray}\label{hkf8}
H_{8}(k)=\left(\begin{array}{ccc}
-\zeta_2(k) & -\zeta_1(k) &-\zeta_0(k)\\
1 & 0 & 0\\
0 & 1 & 0
\end{array}
\right).
\end{eqnarray}
The eigenbands of $H_{8}(k)$ are plotted in Fig. \ref{f8}(f). We can clearly see the three band strings form a figure-8 knot, which is isotopic to the braid diagram in Fig. \ref{f8}(a).\\

\noindent\underline{\textit{\textbf{Example 2: Whitehead link}}}
Similarly, we can work out the NH Hamiltonian $H_{w}(k)$ for the Whitehead link. The braid diagram $B_K$ (see Fig. 1 in the main text) is described by braid word $\tau_1\tau_2^{-1}\tau_1\tau_2^{-1}\tau_2^{-1}$, with total crossing number $c[K]=5$. The permutation associated with $B_K$ is $\sigma=\left(\begin{array}{cccc}
1 & 2 & 3\\
3 & 2 & 1
\end{array}\right)$. There are two cycles $\sigma=(13)(2)$ in $\sigma$. $s_1=(13)$ with length $l_{1}=2$ and $s_2=(2)$ with length $l_{2}=1$. 

To parameterize the two cycles, we need to identify all the data points of trigonometric interpolation. For cycle $s_1$, we pick $c[K]l_{1}=10$ evenly distributed points along $k$ direction, with coordinates $(0,-1)$, $(\frac{\pi}{5},0)$, $(\frac{2\pi}{5},1)$, $(\frac{3\pi}{5},1)$, $(\frac{4\pi}{5},0)$, $(\pi,1)$, $(\frac{6\pi}{5},1)$, $(\frac{7\pi}{5},0)$, $(\frac{8\pi}{5},-1)$, $(\frac{9\pi}{5},-1)$ on the $xz$ plane. For cycle $s_2$, we pick $c[K]l_{2}=5$ evenly distributed points with coordinates $(0,0)$, $(\frac{2\pi}{5},-1)$, $(\frac{4\pi}{5},-1)$, $(\frac{6\pi}{5},0)$, $(\frac{8\pi}{5},1)$. The discrete Fourier transformation yields
\begin{eqnarray}
F_{1}(k)&=&0.1-0.79\cos k+0.57\sin k-0.16\cos 2k+0.5\sin 2k-0.11\cos 3k-0.35\sin 3k\notag\\
&&+0.06\cos 4k+0.04\sin 4k-0.1\cos 5k,\notag\\
F_{2}(k)&=&-0.2+0.32\cos k-\sin k-0.12\cos 2k-0.09\sin 2k.
\end{eqnarray}
To obtain $G_{1}(k)$ and $G_{2}(k)$, we solve all the crossings of the three strings $F_{1}(\frac{k}{2})$, $F_{1}(\frac{k+2\pi}{2})$, and $F_{2}(k)$ inside $[0,2\pi]$. The solution of $F_{1}(\frac{k}{2})=F_{1}(\frac{k+2\pi}{2})$ is $k_p=1.8850$; The solutions of $F_{1}(\frac{k}{2})=F_{2}(k)$ are $k_p=0.6004,4.23291,5.8202$; The solution of $F_{1}(\frac{k+2\pi}{2})=F_{2}(k)$ is $k_p=3.1696$. The interpolation data for $G_{1}(k)$ and $G_{2}(k)$ on the $yz$ plane are respectively $(\frac{0.6004}{2},-1)$, $(\frac{1.8850}{2},1)$, $(\frac{4.2329}{2},-1)$, $(\frac{5.8208}{2},1)$, $(\frac{1.8850+2\pi}{2},-1)$, $(\frac{3.1696+2\pi}{2},1)$ and $(0.6004,1)$, $(3.1695,-1)$, $(4.2329,1)$, $(5.8202,-1)$. The trigonometric interpolation reads
\begin{eqnarray}
G_{1}(k)&=&0.26+0.11\cos k-0.40\sin k-0.27\cos 2k-0.37\sin 2k-1.32\cos 3k,\notag\\
G_{2}(k)&=&1.03-0.12\cos k+1.47\sin k-2.11\cos 2k.
\end{eqnarray}
According to Eq. (\ref{che}), the ChP is
\begin{eqnarray}
f(\lambda,k)&=&[\lambda-F_{2}(k)-iG_{2}(k)]\prod_{j=0}^1[\lambda-F_{1}(\frac{k+2\pi j}{2})-iG_{1}(\frac{k+2\pi j}{2})]\notag\\
&=&\lambda^3+\zeta_2(k)\lambda^2+\zeta_1(k)\lambda+\zeta_0(k).\notag\\
\end{eqnarray}
where
\begin{eqnarray}
\zeta_2(k)&=&-1.56i+0.66i\cos k-0.74i\sin k+2.11\cos 2k,\notag\\
\zeta_1(k)&=&(-0.25+1.07i)-(0.01+1.22i)\cos k+(0.34+1.04i)\sin k+(0.36-2.14i)\cos 2k+(0.73+0.64i)\sin 2k\notag\\&&+(0.37+0.13i)\cos 3k-(0.83+1.44i)\sin 3k-(0.01+0.26i)\cos 4k-(0.04+0.09i)\sin 4k,\notag\\
\zeta_0(k)&=&(0.19-1.05i)+(0.77+0.76i)\cos k-(0.07+0.54i)\sin k-(0.40-1.61i)\cos 2k-(0.64-0.23i)\sin 2k\notag\\&&+(1.26-1.28i)\cos 3k-(0.37+0.39i)\sin 3k+(0.73-0.33i)\cos 4k-(0.47+0.44i)\sin 4k\notag\\
&&+(0.22+0.88i)\cos 5k+(0.51-0.06i)\sin 5k+(0.14-0.02i)\cos 6k-0.04i\sin 6k-0.01i\cos 7k.
\end{eqnarray} 
The NH Hamiltonian $H_{w}(k)$ follows from Eq. (\ref{hk}) with $\zeta_j(k)~(j=0,1,2)$ listed above.

\subsection{(V) Revealing knots from quench dynamics}
Here we provide more details about how to reveal the various knots [See Fig. 2(b) in the main text] from the quench dynamics of two-band systems. Due to the complex eigenenergies, the time evolution of a NH system is non-unitary. Let us suppose the initial state at time $t=0$ to be $|\xi_0\rangle$ and denote the $j$-th eigenenergy of the NH Hamiltonian to be $E_j(k)=\epsilon_j-i\gamma_j$. At a later time $t$, the time-evolved state is ($\hbar=1$)
\begin{eqnarray}
|\xi(k,t)\rangle=\sum_{j} e^{-i\epsilon_jt}e^{-\gamma_j t}\langle\chi_{j}|\xi_0\rangle|\psi_{j}\rangle,
\end{eqnarray}
where $|\psi_j\rangle$ and $|\chi_j\rangle$ are the right and left eigenvectors, respectively.  It is clear that after a long time, the eigenstate with smaller $\gamma_j$ will dominate and the time-evolved state $|\xi(k,t)\rangle$ will fall into the corresponding right eigenstate $|\psi_j\rangle$, in contrary to the Rabi oscillation after quantum quench in Hermitian system. The time-evolved state can be measured through the momentum- and time-resolved Bloch-state tomography \cite{hhpquench,azi1,azi2,tomograph1,tomograph2}.

\begin{figure}[h!]
\includegraphics[width=0.45\textwidth]{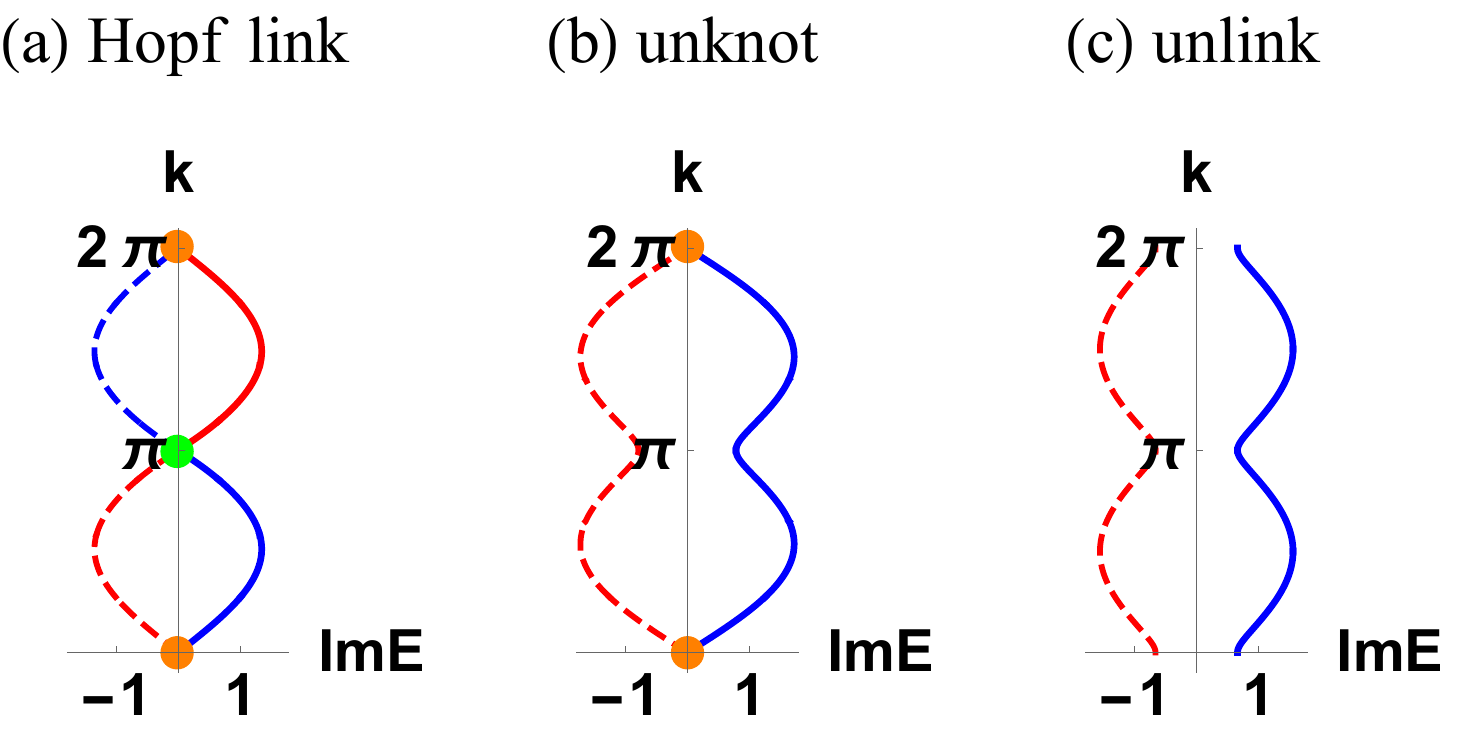}
\caption{Projection of the NH bands of Hamiltonian $H_{12}(k)$ (Eq. (7) in the main text) onto the $(ImE,k)$ plane. (a) The Hopf-link phase, $m_1=m_2=0.5$; (b) The unknot phase, $m_1=m_2=0.9$; (c) The unlink phase, $m_1=1.2$, $m_2=0$. The green/orange dots label the $k=0/\pi$ mode where the imaginary parts of the two eigenvalues are equal. The solid (dotted) curves represent the NH bands with $ImE>0$ ($ImE<0$). The long-time quench dynamics selects the eigenstates with $ImE>0$ represented by solid curves.}\label{figs2}
\end{figure}

We note that different knots, i.e., different braiding patterns of eigenenergy strings, have different projections onto the plane spanned by $k$ and the imaginary energy $ImE$. For the three knots shown in Fig. 2(b) in the main text, the corresponding projections of the eigenenergy strings on the $(ImE, k)$ plane are depicted in Fig. \ref{figs2}(a)(b)(c). The quench dynamics then selects the eigenbands with $ImE>0$ (solid curves). For the Hopf-link phase [Fig. \ref{figs2}(a)], the projection contains both bands, connected by the two overlapping points at $k=0$ and $k=\pi$, where the imaginary parts of two eigenvalues are equal. For the unknot phase [Fig. \ref{figs2}(b)], the projection has one overlapping point at $k=0$. Note that the bands at $k=0$ and $k=2\pi$ are interchanged due to band braiding. While for the unlink phase [Fig. \ref{figs2}(c)], the projection consists of only one single band.

It may seem that by going from $(ReE, ImE, k)$ space to $(ImE, k)$ space, we only get to see the “shadow” of the knots. However, in practice and especially for simple knots, the partial information is enough to determine the knot structure. The two eigenvectors of $H_{12}(k)$ form closed loops on the Bloch sphere. The knot topology of two eigenenergy strings translates to characteristic crossing patterns of the loops on the sphere. As illustrated in Fig. 3 in the main text, the Hopf link gives two loops intersecting twice with each other; the unknot gives one closed loop; while the unlink gives two separated loops. These distinctions unambiguously identify the underlying knot structure. The band permutation and invariant $Q$ can be read out directly. The long-time evolution of the three phases exhibits different behaviors, as illustrated in Fig. 3(a)(b)(c) in the main text. For the Hopf-link phase, the $k\in(0,\pi)$ and $k\in(\pi,2\pi)$ modes come from different bands. The $k = 0$ and $k=\pi$ modes are special, whose postquench states precess along a big circle on the Bloch sphere due to their equal imaginary part of eigenenergies. For the unknot phase, the long-time dynamics falls into an open arc on the Bloch sphere, except for the precessing $k=0$ mode. For the unlink phase, the long-time evolution of all $k$-modes falls into one of two closed curves.

\subsection{(VI) Proposal for experimental realization of Hamiltonian $H_{12}$ in electric circuits and photonic arrays}

In this section, we detail how to experimentally realize the NH Hamiltonian $H_{12}$ in electric circuits and photonic arrays. \\

\noindent\underline{\textit{\textbf{Realization in RLC circuits}}} Circuits with resistor, inductor, and capacitor (RLC) components constitute a powerful platform for simulating NH systems. The circuit Laplacian, which governs the behavior of the RLC circuits, play a similar role to the Hamiltonian in a quantum system. In this section, we detail how to realize the two-band Hamiltonian (Eq. (7) in the main text) using the RLC circuits. The circuit is dictated by the Kirchhoff's law, 
\begin{eqnarray}\label{kirlaw}
I_i=\sum_{j}g_{ij}(V_i-V_j)+w_i V_i,
\end{eqnarray}
where $I_i$ and $V_i$ are the input current and voltage at the $i$-th node. The first term in Eq. (\ref{kirlaw}) is the current flowing towards all other nodes $j$ from the $i$-th node linked by conductance $g_{ij}$; the second term is the current flowing into the ground with conductance $w_i$. Denote the current and voltage vectors as $\bm I=(I_1,I_2,...)$ and $\bm V=(V_1,V_2,...)$, Eq. (\ref{kirlaw}) is recast into the matrix form \cite{RLC3}
\begin{eqnarray}
\bm I=(D-G+W)\bm V\equiv J\bm V,
\end{eqnarray}
where $D=diag(\sum_{j}g_{1j},\sum_{j}g_{2j},...,)$ lists the total conductance out of each node; $G=(g_{ij})$ lists the conductance between different nodes; and $W=diag(w_1,w_2,...)$ lists the grounded conductance. $J=D-G+W$ is the Laplacian (or admittance matrix). In RLC circuits, the conductance is real for resistive elements and complex for capacitors or inductors.

\begin{figure}[h!]
\includegraphics[width=0.78\textwidth]{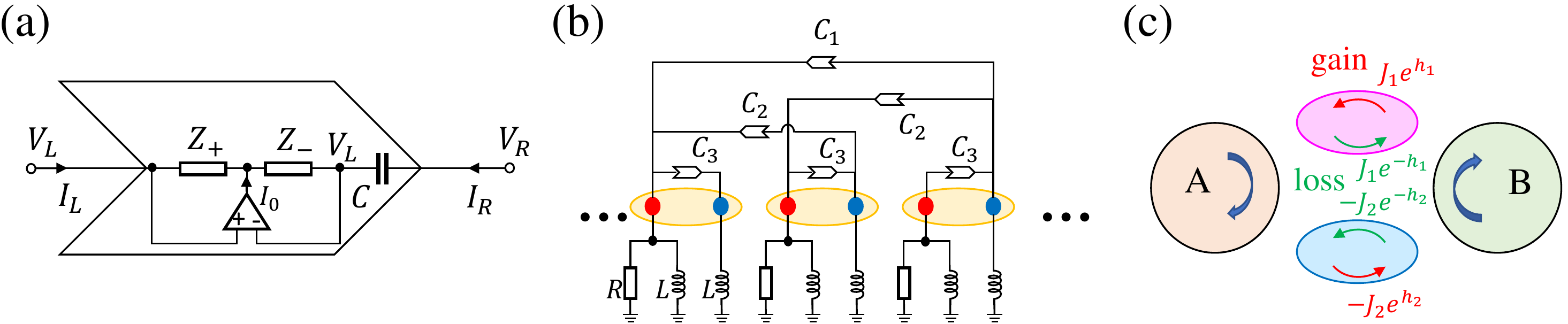}
\caption{(a) The configuration of INIC, which consists of an operation amplifier, positive (negative) feedback impedance $Z_{+}$ ($Z_{-}$) and a capacitor. (b) Schematics of the RLC circuits to realize our NH Hamiltonian (Eq. (7) in the main text). Each unit cell contains two sites, represented by the red and blue nodes. The periodic boundary condition is used. Only part of the periodic lattice is shown. (c) Schematics of realizing asymmetric (unidirectional) coupling between the two main cavities $A$, $B$ (orange and green) via two auxiliary microring cavities (magenta and blue). The red/green arrows denote light amplification/attenuation in the upper or lower half perimeter of the microring cavities.}\label{figs3}
\end{figure}

To realize the non-reciprocal couplings in the NH Hamiltonian, one can use the impedance converter with current inversion (INIC), which consists several RLC components and an operation amplifier \cite{nhsee2}, as shown in Fig. \ref{figs3}(a). Based on the negative feedback configuration, the current and voltage relation is \cite{nhsee2}
\begin{eqnarray}
\left(\begin{array}{cc}
I_L\\
I_R\end{array}\right)=i\omega C\left(\begin{array}{cc}
-\frac{Z_-}{Z_+} & \frac{Z_-}{Z_+}\\
-1 & 1\end{array}\right)
\left(\begin{array}{cc}
V_L\\
V_R\end{array}\right),
\end{eqnarray}
where $Z_{\pm}$ are the total impedance of the positive/negative feedback loops. Note that $I_L\neq I_R$ because the operation amplifier is an active circuit element and acts as a current source or drain. In the simple case of $Z_-=0$, the relation simplifies, and the INIC is completely described by the capacitance $C$.

Fig. \ref{figs3}(b) depicts a setup of RLC circuits to implement the NH Hamiltonian $H_{12}$ (Eq. (7) in the main text). It contains non-reciprocal coupling terms, provided by the above INIC components. We choose the simple case with $Z_-=0$. The circuits work in the linear regime, where the complex admittance for the capacitors, resistors and inductors are given by $i\omega C$, $1/R$, and $1/i\omega L$, respectively. Each unit cell in the 1D lattice contains two nodes (red and blue). The intra-cell, nearest neighbor, and next nearest neighbor couplings are given by three INICs characterized by capacitance $C_3$, $C_2$, and $C_1$ respectively. The current-voltage relation can be easily written as
\begin{eqnarray}
\left(\begin{array}{cc}
I_{i,red}\\
I_{i,blue}\end{array}\right)=\left(\begin{array}{cc}
(i\omega C_1+i\omega C_2+\frac{1}{R}+\frac{1}{i\omega L})\delta_{i,j} & -i\omega C_1\delta_{i+2,j}-i\omega C_2\delta_{i+1,j}\\
-i\omega C_3\delta_{i,j} & (i\omega C_3+\frac{1}{i\omega L})\delta_{i,j} \end{array}\right)\left(\begin{array}{cc}
V_{j,red}\\
V_{j,blue}\end{array}\right).
\end{eqnarray}
Under periodic boundary condition, the circuit Laplacian is Fourier transformed into 
\begin{eqnarray}
J(\omega)=-i\omega C_1\left(\begin{array}{cc}
-1-\frac{C_2}{C_1}+\frac{1}{\omega^2 C_1L}+\frac{i}{\omega C_1R}~~&~~ e^{i 2k}+\frac{C_2}{C_1}e^{i k}\\
\frac{C_3}{C_1}~~ &~~ -\frac{C_3}{C_1}+\frac{1}{\omega^2 C_1L} \end{array}\right).
\end{eqnarray}
By choosing $C_3=C_1+C_2$, $L=\frac{1}{\omega^2C_3}$ and setting $\frac{C_2}{C_1}=m_2$ and $\frac{1}{2\omega C_1 R}=m_1$, we finally arrive at the NH $H_{12}(k)$ in the main text after neglecting a constant term $-i m_1\sigma_0$. 

In this way, we have established a one-to-one correspondence between the NH Hamiltonian and the admittance matrix. Upon setting up the circuits, the band structure and eigenstates can be directly measured by the admittance and voltage measurements \cite{RLC3,nhsee2,RLCmeasure1,RLCmeasure2}. For our periodical boundary case, by respectively feeding an input current to the two different sublattices and measuring the voltage at all nodes, the admittance spectrum is then reconstructed through a Fourier transformation. In particular, the exceptional points should reveal themselves through divergent impedance known as topolectrical resonance \cite{RLC3}. More complicated NH Hamiltonians such as $H_8$ and $H_w$ can be realized by properly choosing the unit cells and their couplings along the lines presented here. \\

\noindent\underline{\textit{\textbf{Realization in photonic arrays}}} Next, we turn to the experimental realization in photonic systems with coupled cavity arrays. Similar to the above proposal in electric circuits, the key ingredient is to realize the asymmetric (unidirectional) coupling between the two sublattices (denoted as $A$, $B$), i.e., the NH term $\sim\left(\begin{array}{cc}0 & 1\\ 0 & 0\end{array}\right)$. The NH Hamiltonian $H_{12}$ can then be obtained by connecting different lattice sites with this asymmetric coupling. In the following, we address how to realize the unidirectional coupling. In the cavity arrays, each lattice site is represented by a main cavity, which is indirectly coupled to its neighbor via two auxiliary microring cavities, as shown in Fig. \ref{figs3}(c). The two microring cavities provide light amplification and attenuation in the upper or lower half perimeter \cite{longhi}. We note that the coupling amplitude and phase, and the photon gain and loss for each cavity, can be controlled independently \cite{cavity1,op14}. The net tunnelings from $A$ to $B$ and $B$ to $A$ are $J_1e^{-h_1}-J_2e^{h_2}$ and $J_1e^{h_1}-J_2e^{-h_2}$, respectively. By choosing $J_1e^{-h_1}-J_2e^{h_2}=0$, the two tunneling channels from $A$ to $B$ cancel each other. Only the tunnelings from $B$ to $A$ survive, thus realizing the desired unidirectional coupling. 

\subsection{(VII) NH knot and skin effect}

In the main text, we have discussed several observable consequences of the knot topology, e.g., the emergence of EPs between different knotted phases and the exotic quench dynamics. These features are associated with the NH Bloch bands, i.e., when periodic boundary condition (PBC) is taken. Within the broader context of NH physics, it remains an open question to relate the knot topology with other physical consequences, especially those under open boundary condition (OBC). We note the intrinsic differences between the NH band topology and the familiar Hermitian cases, due to the complex eigenenergy in NH systems. A typical example is the NH skin effect, where extensive number of eigenstates are localized at the boundary and certain physical quantities are sensitive to the boundary conditions. The sensitivity with boundary conditions obscures the standard diagnostic toolset for Hermitian band topology, like bulk-boundary correspondence and quantized response. 

\begin{figure}[h!]
\includegraphics[width=0.62\textwidth]{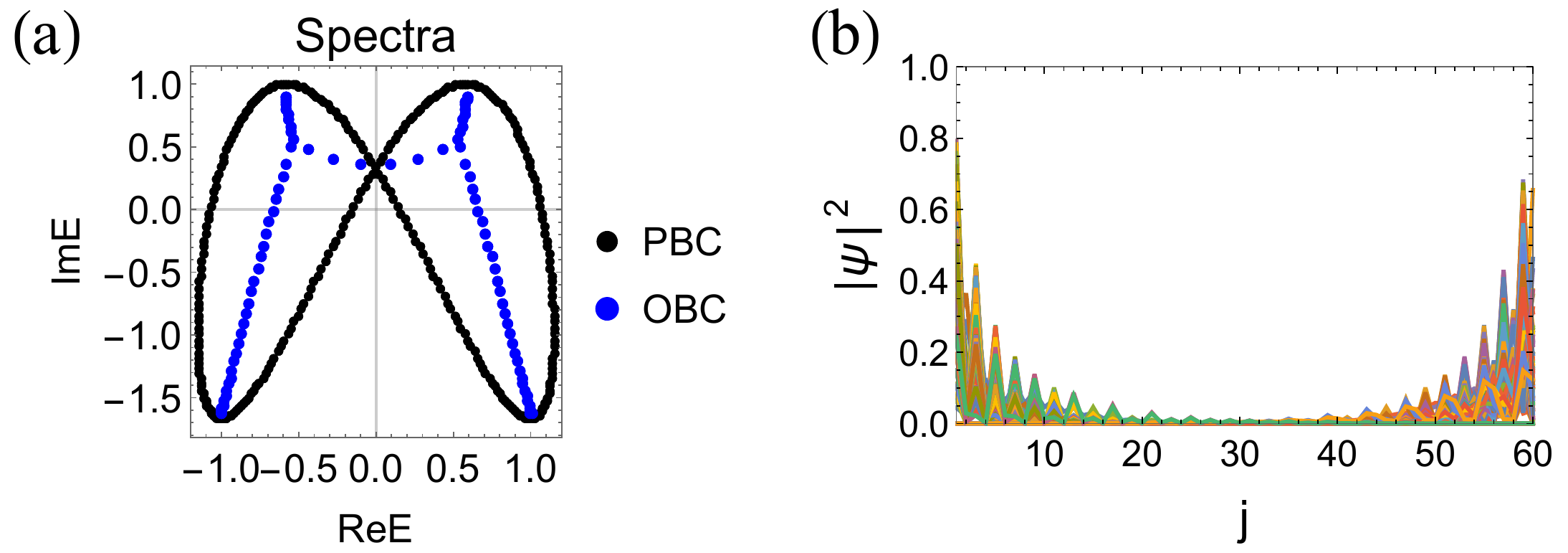}
\caption{(a) Energy spectra of Hamiltonian (\ref{hkf8}) under different boundary conditions on the complex plane. The Bloch bands tie into a figure-8 knot (see Fig. \ref{f8}(f)). The projection on the complex plane forms a closed loop (black). The spectra under OBC (blue) collapse into the interior region of the loop. The number of lattice site is $L=60$. (b) Spatial profiles of all the eigenstates under OBC, showing the NH skin effect. Here $j$ is the site index along the 1D chain.}\label{figs4}
\end{figure}

The knotted phases inevitably exhibit skin effect as projecting the knot onto the complex energy plane yields a band structure with a point gap \cite{pointtopo1,pointtopo2,pointtopo3}. As an example, we consider the figure-8 knot, described by Hamiltonian (\ref{hkf8}). The spectra under different boundary conditions are depicted in Fig. \ref{figs4}(a). On the complex energy plane, the spectra under PBC form a closed loop and are the projection of the figure-8 knot in Fig. \ref{f8}(f). While under OBC, the spectra collapse into the interior region of the loop. The wave functions of the OBC eigenstates are depicted in Fig. \ref{figs4}(b). These eigenmodes are fully localized at the two ends of the chain, manifesting the NH skin effect. Further information of the skin modes can be revealed by introducing the concept of generalized Brillouin zone \cite{ne1}.

\begin{thebibliography}{99}
\bibitem{fnote1} Finding the conjugacy classes of $B_N$ is hard problem, see Refs. \cite{class2,knotbook}.
\bibitem{fnote2} The name twister here refers to the phase twist, and the fact that the eigenenergy strings (see $T_2$, $T_3$ in Fig. 1) resemble a tornado or twister. It should not be confused with twistor as in twistor theory.
\end{thebibliography}
\end{document}